# International students' loneliness, depression and stress levels in COVID-19 crisis. The role of social media and the host university

Nikolaos Misirlis[1], Miriam H. Zwaan[1], David Weber[2]

1. HAN University of Applied Sciences, International School of Business, Arnhem, The Netherlands

2. University of North Carolina Wilmington, USA

## Abstract

The move to university life is characterized by strong emotions, some of them negative, such as loneliness, anxiety, and depression. These negative emotions are strengthened due to the obligatory lockdown due to the COVID-19 pandemic. Previous research indicates association among the use of social media, university satisfaction, and the aforementioned emotions. We report findings from 248 international undergraduates in The Netherlands, all students at the International School of Business. Our results indicate strong correlations between anxiety, loneliness, and COVID-19-related stress with university satisfaction together with social capital.

**Keywords:** COVID-19; Pandemic; lockdown; loneliness; depression; anxiety; international students

## 1. Introduction

### 1.1. Transition to University life abroad

The transition from secondary school to a university and its academic challenge is admittedly one of the most important moments in the development of young people (Thomas, Orme, & Kerrigan, 2020). In cases when the student decides to study abroad, the transition may be that much harsher and more abrupt (Jackson, 2003; Rokach, 1989). Friendships or family relationships are entering a different phase. In some cases, the transition from school to academia is more abrupt and harsh, as students decide to study abroad.  In this study, we are focusing in particular on students we refer to as international or internationally mobile, which we define as students who have crossed a national or territorial border for the purpose of education and are now enrolled in institution outside their country of origin (Unesco). There is a dramatic increase in international students enrolled in Dutch universities during the last decade. In 2018-2019 more than 85 thousand international students were studying in The Netherlands, representing an 11.5% of the total number of students enrolled (nuffic.nl, 2019).

Expatriating and living abroad is accompanied by objective difficulties at all ages. Even more at sensitive ages where psychological pressure may be greater as life experience is less (Hunt & Eisenberg, 2010).

### 1.2. Psychological aspects of moving to university

In addition, this psychological pressure can be greater, as in most cases, the transition to student life is individual, without the accompaniment of a friend or family member. That is why most universities, recognize the importance of Introduction Weeks (Gale & Parker, 2014; Sullivan & Kashubeck-West, 2015). Academic staff, along with psychologists and other professionals, accompany and adapt students to the new reality, thus helping to make it easier for young people to acclimatize and comfort the transition shock and stress (Berry, 1997; Li, Marbley, Bradley, & Lan, 2016; Lin, 2006; Lowinger, He, Lin, & Chang, 2014).

Many times this transition is accompanied by stressful situations. Emotions become more intense when students feel - or actually are – alone (Cacioppo, Hughes, Waite, Hawkley, & Thisted, 2006; Richardson, Abraham, & Bond, 2012). Unfortunately, in many cases, loneliness combined with anxiety can lead to unpleasant results.

### 1.2. The pandemic overturn and the COVID-19 reality

Today, stressful situations and loneliness due to the pandemic are even more intense. In particular, in international universities where student families are often thousands of miles away in combination with compulsory lockdown due to the pandemic, they create an explosive mix of emotions that often leads to increased stress or even depression. The ongoing COVID-19 pandemic was confirmed to have spread to the Netherlands on 27 February 2020, when its first COVID-19 case was confirmed. On Thursday March the 12$^{th}$, the Dutch government announced an intelligent lockdown and prohibited schools and universities to continue with face to face education.

Research into the COVID-19 pandemic's extensive impact - in university life in particular- is currently in the earliest stages of development. Researchers in several branches of the physical and social sciences are attempting to place pebbles small or large in what will eventually become an ocean of knowledge. The present study is one such attempt, and is designed to trace the psychological problems caused by the COVID-19 issue and will discover correlations between the various emotions that an international student feels and that may be reinforced, caused or increased by the lockdown specifically in the lived experience of international students.

## 2. Method

### 2.1. Participants

We collected data from a medium-size public university in the south central region of the Netherlands. The research was focused on students belonging to the two international streams of the ISB, Communication Studies (CS) and International Business (IB). Students in this specific institution study for four years, starting with a propaedeutic year. In their third year, they study one semester abroad at one of our partner universities worldwide followed by a mandatory internship of one semester. IB students obtain a BSc degree, while CS students obtain a BA degree. Due to the lockdown, surveys were only distributed online for a period of exactly one month (from 6$^{th}$ of April to 5$^{th}$ of May). In a total of 248 valid responses, 171 were female students and 71 males, 99 students study to the CS stream, 144 in IB and 5 students are from exchange programs (i.e. Erasmus). 105 are freshmen, 94 in their second year of study, 27 at the third, 20 at the fourth and 2 students graduated last summer. The dominant nationality is the Dutch (n=62) and Vietnamese (n=37). We included student of Dutch nationality for two main reasons. One, the culture and interactional environment at the university is, by design and circumstance, highly internationalized. The lingua franca -- the primary working language -- of the university is English, not Dutch. Two, at this university, a student with Dutch nationality would typically have been born outside The Netherlands, likely in a former Dutch colony that before his or her birth had become an independent nation (e.g., Aruba, St. Maarten). Such a student tends to have dual citizenship, and therefore can claim two nationalities; and has often lived abroad for many years.

### 2.2 Survey design

We compiled a 98-item survey on ThesisToolPro application. The survey remained active to accept responses for one month exactly, starting from April the 6$^{th}$ (reminder: the university facilities closed due to the pandemic on the 13$^{th}$ of March). We collected information about demographics, social media usage such as habits, intensity, measurable activities such as the number of friends the university satisfaction of the students, their levels of loneliness, anxiety, and depression as well as the psychological sense of belonging to the university community. Bridging and bonding social capital were also measured, based on published scales. Scale items, survey and relevant information can be found in Appendix A).

### 2.3 Scales and measures/ Internal consistency scores

*Loneliness*

In order to measure students' loneliness, we used the UCLA Loneliness scale (Russell, Peplau, & Cutrona, 1980) (see Appendix A). The Cronbach alpha for the 20-item scale was 0.9261.

*Social capital*

We used Ellison, Steinfield, and Lampe (2007) scales in order to measure social capital. This scale contains items regarding the bridging the social capital and bonding social capital. The Cronbach alpha for this scale was 0.8789.

*University satisfaction*

To measure the students' experience and satisfaction from the institution we used the Three-Factor Psychological Sense of Community scale of Jason, Stevens, and Ram (2015). Current literature indicates that the first weeks of the new academic life of students are crucial and new university plays a very significant role in that by welcoming, facilitating and informing the new students (Lemma, Gelaye, Berhane, Worku, & Williams, 2012; Richardson et al., 2012). The Cronbach alpha for the University satisfaction was 0.8588.

*Social media use*

In order to measure the use of social media among students, we used the social media intensity scale from Ellison et al. (2007), the bounded self scale from Boyd and Ellison (2007) and the liminal self scale from Kerrigan and Hart (2016). Cronbach alpha for this scale was 0.7822.

*Anxiety and depression*

To measure the anxiety and the depression we used the Hospital Anxiety and Depression scale (HADS; Snaith and Zigmond (2000)). The 14-item scale was divided into two 7-items scales, one for anxiety and one for depression, obtaining a Cronbach alpha equal to 0.82 and 0.7969 respectively. The combined Cronbach alpha was 0.8604.

*COVID-19*

The pandemic-related scale was a rather original set of items we used based on previous research (of course not related to COVID, since this is the first pandemic situation since 1918 and the Spanish flu). In order to measure different aspects of the lockdown due to COVID-19 in students' life, we adapted and borrowed items from the General Anxiety Disorder-7 (GAD-7; Spitzer, Kroenke, Williams, and Löwe (2006)), the Center of Epidemiologic Studies Depression (LES-D; Eaton, Smith, Ybarra, Muntaner, and Tien (2004); Radloff (1991); Roberts, Andrews, Lewinsohn, and Hops (1990)) and the Impact of Event Scale-Revised (IES-R; Creamer, Bell, and Failla (2003); Weiss (2007)) scale obtaining a Cronbach alpha equal to 0.9046.

## Results

We conducted bivariate correlation analysis, obtaining some statistically significant results (Table 1). The stress, panic and anxiety from the pandemic lockdown are associated with high levels of loneliness (c: 0.339; p: 0.01). As expected, depression is also associated with high levels of loneliness (c: 0.410; p: 0.01). The university satisfaction is also associated with low levels of depression (c: -0.217; p: 0.01). The lockdown experience on a student's life is correlated to high levels of anxiety (c: 0.710, p:0.01). This index, together with the correlation between University satisfaction and social capital (c: 0.737; p: 0.01), represents the highest percentage of predictions (71% and 73.7% respectively).

*Table 1: Bivariate correlations and descriptive statistics for all variables (N = 248).*

| | | 1 | 2 | 3 | 4 | 5 | 6 | 7 |
|---|---|---|---|---|---|---|---|---|
| 1 | **Loneliness** | 1 | | | | | | |
| 2 | **COVID-19** | .339** | 1 | | | | | |
| 3 | **Anxiety** | .279** | .710** | 1 | | | | |
| 4 | **Depression** | .410** | .319** | .238** | 1 | | | |
| 5 | **University Satisfaction** | .401** | -.164** | -.128* | -.217** | 1 | | |
| 6 | **Social Capital** | -.535** | -.159* | -.116 | -.222** | .737** | 1 | |
| 7 | **Social media use** | -.116 | .112 | .078 | -.156* | .273** | .311** | 1 |

**. Correlation is significant at the 0.01 level (2-tailed).

*. Correlation is significant at the 0.05 level (2-tailed).

Regarding the COVID-19-related items, the is an apparent association, though weak, between COVID-19 and social capital (c: -0.159; p: 0.05) and COVID-19 and University satisfaction (c: - 0.164; p: 0.05). A possible explanation for these low indexes may be that the survey was distributed after 23 days of lockdown where the members of the academic community had only just begun to manage the massively unsettling changes in cognitive, emotional, and operational functioning brought about by the surging pandemic and sudden lockdown. Many potential respondents were benumbed, and therefore not particularly interested in completing surveys

## 4. Discussion

Our research interest revolves around the correlations among the psychological aspects of international students, their satisfaction over the university, their social capital and the use of the most dominant social media platforms. Especially for the psychological aspects, we focused on common issues that students may confront such as anxiety, loneliness and depression. Furthermore, the lockdown due to the coronavirus pandemic crisis brought to the fore stress issues that the students may feel, especially those living thousands of miles away from their countries and therefore their comfort zones. Universities, as suggested by the results of this study, can be one of the most crucial inhibitors for this category of stress or anxiety. Loneliness, anxiety, or depression – or lately, the COVID-19-related stress psychological aspects – can be seriously lowered down if the host institutions provide moral and psychological support to the students. Together with that, social capital, digital friends and university colleagues can be beneficial.

*What Universities can do*

Surely no institution could have predicted, or been fully prepared for, a global crisis of the coronavirus pandemic's magnitude as it emerged in March and April of 2020. The most recent pandemic of this scope occurred over one hundred years earlier, a world that few people alive today can recall, characterized by sociocultural features almost wholly different from our own. Social media have completely changed the way we act, react, study, communicate and solve problems. Today, the entire scientific community tries to find a solution to this pandemic. Universities have now the opportunity to create knowledge for the future. In our case, social science universities and international schools can create infrastructures for possible future similar crises. Our research indicates a strong connection with university satisfaction and the students' well-being. Universities must move towards that direction and build strong relationships with the students, in our case the international students' community. During this crisis we learned – and we still keep learning – to appreciate collateral values, other than the strict knowledge. We, as academics, must listen to our students and create high standards regarding the connection between them and the institution.

**5. Limitations – Future implications**

The problems due to the pandemic lockdown are related to time, sources and the necessity of quick publication. We had to run the survey only for a one-month-time span sending only one reminder to the students. This is the reason that we obtained only a 21.1% response rate (248 responses from 1150 emails). Many of the students have returned to their countries, having limited access to the institution's account. Furthermore, since the community needs quick results from COVID-19-related research, we decided not to wait for a second round of responses. The aforementioned limitations, though, can lead to future research, covering these restrictions. Moreover, the current literature is limited and therefore, a systematic review will be necessary and useful, in a future moment, when there will be a plethora of relevant articles. Localization of research is another limitation we face. Further research to other international schools (outside The Netherlands) and comparison analyses can be beneficial for science.

Finally, the present study will lead to practical implications, such as the formation of Universities Pandemic Crisis teams all over the world with common manuals and good practices, exchanging experience from research like the present.

## Appendix

| **Scale** | **Items** * = reverse scored |
|---|---|
| Demographics | Age<br>Year of study<br>Place of birth<br>Nationality<br>Gender<br>CS/IB/ Exchange student |
| Revised UCLA Loneliness Scale | I feel in tune with the people around me.*<br>I lack companionship.<br>There is no one I can turn to.<br>I do not feel alone.*<br>I feel part of a group of friends.*<br>I have a lot in common with the people around me.*<br>I am no longer close to anyone.<br>My interests and ideas are not shared by those around me.<br>I am an outgoing person.*<br>There are people I feel close to.*<br>I feel left out.<br>My social relationships are superficial.<br>No one really knows me well.<br>I feel isolated from others.<br>I can find companionship when I want it.*<br>There are people who really understand me.* |
| | I am unhappy being so withdrawn.<br><br>People are around me but not with me.<br><br>There are people I can talk to.*<br><br>There are people I can turn to.* |
| Depression and anxiety scale (HADS scale) | **Items of anxiety**<br><br>I feel tense or wound up<br><br>I get a sort of frightened feeling as if something bad is about to happen<br><br>Worrying thoughts go through my mind<br><br>I can sit at ease and feel relaxed*<br><br>I get a sort of frightened feeling like butterflies in the stomach<br><br>I feel restless and have to be on the move<br><br>I get sudden feelings of panic<br><br>**Items of depression***<br><br>I still enjoy the things I used to enjoy<br><br>I can laugh and see the funny side of things<br><br>I feel cheerful<br><br>I feel as if I am slowed down<br><br>I have lost interest in my appearance<br><br>I look forward with enjoyment to things<br><br>I can enjoy a good book or radio or TV programme |
| University satisfaction – Introduction from the institution | The introduction I received was helpful.<br><br>I received a welcome pack to introduce me to the area.<br><br>I had a positive experience meeting my hall of residence/flatmates for the first time. |
| | I felt knew who my course mates were.<br><br>I don't feel I got enough support from my university.*<br><br>I knew how to contact my course mates.<br><br>More could be done to help support students when they first move to university.*<br><br>I had a positive experience moving to university. |
| University satisfaction – Psychological sense of community scale | I think this university is a good university.<br><br>I am not planning on leaving this university.<br><br>For me, this university is a good fit.<br><br>Students can depend on each other in this university.<br><br>Students can get help from other students if they need it.<br><br>Students are secure in sharing opinions or asking for advice.<br><br>This university is important to me.<br><br>I have friends in this university.<br><br>I feel good helping the university and the students. |
| Social Capital –<br><br>Bridging social capital | I feel I am part of the university community.<br><br>I am interested in what goes on at my university.<br><br>My university is a good place to be.<br><br>I would be willing to contribute money to my university after graduation.<br><br>Interacting with people at my university makes me want to try new things.<br><br>Interacting with people at my university makes me feel like a part of a larger community.<br><br>I am willing to spend time to support general university activities.<br><br>At my university I come into contact with new people all the time.<br><br>Interacting with people at my university reminds me that everyone in the world is connected. |
| Social Capital –<br><br>Bonding social capital | There are several people at my university I trust to solve my problems.<br><br>If I needed an emergency loan of 100€, I know someone at my university I can turn to.<br><br>There is someone at my university I can turn to for advice about making very important decisions.<br><br>The people I interact with at my university would provide good job references for me.<br><br>I do not know people at my university well enough to get them to do anything important.* |
| Social media intensity scale | About how many total _______ friends do you have? 0 = 10 or less, 1 = 11–50, 2 = 51–100, 3 = 101–150, 4 = 151–200, 5 = 201–250, 6 = 251–300, 7 = 301–400, 8 = more than 400 |
| | In the past week, on average, approximately how many minutes per day have you spent on _______? 0 = less than 10, 1 = 10–30, 2 = 31–60, 3 = 1–2 hours, 4 = 2–3 hours, 5 = more than 3 hours |
| | I am proud to tell people I'm on Facebook/ Instagram/ other<br><br>Facebook/ Instagram/ other has become part of my daily routine.<br><br>I feel out of touch when I haven't logged onto Facebook/ Instagram/ other for a while.<br><br>I feel I am part of the Facebook/ Instagram/ other community.<br><br>I would be sorry if Facebook/ Instagram/ other shut down. |
| Social Media Use - Bounded self | I do not think twice about posting personal data on social media sites.<br><br>I post what I am doing immediately and then forget about it.<br><br>I have no problem with people I do not know seeing my social media data.<br><br>I do not worry if I get a negative reaction to my social media posts.<br><br>I am completely comfortable with being open about myself on social media. |
| Social Media Use - Liminal self | I like to look back at old posts to see how I have changed.<br><br>I do not like to be reminded of who I used to be by old posts on social media. |
| | When I add new people or new people follow me, I worry about them seeing older posts.<br><br>I like to edit or restrict access to old posts to reflect who I am now.<br><br>I wish I could erase and reinvent my social media identity.<br><br>I view my social media as representing who I really am. |